\documentclass{elsart}
\usepackage{graphicx}
\usepackage{amsfonts}
\usepackage{amssymb}

\newcommand{\hide}[1]{}

\newcommand{\squeezelist}{\setlength{\itemsep}{0pt}}

\newcommand{\notyet}[1]{}

\def\P{{\mathcal P}}

\begin{document}
\begin{frontmatter}

\title{Unfolding Manhattan Towers\thanksref{title}}

\author[md]{Mirela Damian} and
\ead{mirela.damian@villanova.edu}
\author[rf]{Robin Flatland} and
\ead{flatland@siena.edu}
\author[jor]{Joseph O'Rourke\thanksref{nsf}}
\ead{orourke@cs.smith.edu}

\thanks[title]{A preliminary version of this paper appeared in CCCG 2005.}
\thanks[nsf]{Supported by NSF Distinguished Teaching Scholars award
       DUE-0123154.}
\address[md]{Dept. of Computer Science, Villanova University,
Villanova, PA 19085, USA.}
\address[rf]{Dept. of Computer Science, Siena College,
Loudonville, NY 12211, USA.}
\address[jor]{Dept. of Computer Science, Smith College,
Northampton, MA 01063, USA.}

{\makeatletter
 \gdef\xxxmark{%
   \expandafter\ifx\csname @mpargs\endcsname\relax 
     \expandafter\ifx\csname @captype\endcsname\relax 
       \marginpar{xxx}
     \else
       xxx 
     \fi
   \else
     xxx 
   \fi}
 \gdef\xxx{\@ifnextchar[\xxx@lab\xxx@nolab}
 \long\gdef\xxx@lab[#1]#2{{\bf [\xxxmark #2 ---{\sc #1}]}}
 \long\gdef\xxx@nolab#1{{\bf [\xxxmark #1]}}
 \gdef\turnoffxxx{\long\gdef\xxx@lab[##1]##2{}\long\gdef\xxx@nolab##1{}}%
}

\begin{abstract}
We provide an algorithm for unfolding the surface of any orthogonal
polyhedron that falls into a particular shape class we call
Manhattan Towers, to a nonoverlapping planar orthogonal polygon. The
algorithm cuts along edges of a $4 \times 5 \times 1$ refinement of
the vertex grid.
\end{abstract}

\end{frontmatter}

\section{Introduction}
\label{sec:Introduction}
It is a long standing open problem to decide whether the surface of
every convex polyhedron can be \emph{edge unfolded}: cut
along edges and unfolded flat to one piece without
overlap~\cite{do-sfucg-05}. 
It is known that some nonconvex polyhedra have no edge unfolding; a
simple example is a small box sitting on top of a larger box.
However, no example is known of a nonconvex polyhedron that cannot
be unfolded with unrestricted cuts, i.e., cuts that may cross the
interior of faces.

The difficulty of these questions led to the exploration of
\emph{orthogonal polyhedra}, those whose faces meet at right angles.
Progress has been made in two directions: firstly, by restricting
the shapes to subclasses of orthogonal polyhedra, such as the
``orthostacks'' and ``orthotubes'' studied
in~\cite{bddloorw-uscop-98}; and secondly, by generalizing the cuts
beyond edges but with some restrictions.  In particular, a
\emph{grid unfolding} partitions the surface of the polyhedron by
coordinate planes through every vertex, and then restricts cuts to
the resulting grid. The box-on-box example mentioned earlier can be
easily grid unfolded. Recent work on grid unfolding of orthostacks
is reported in~\cite{dm-geuoo-04} and~\cite{dil-gvuo-04}.

Because on the one hand no example is known
of an orthogonal polyhedron that cannot be grid unfolded, and on the
other hand no algorithm is known for grid unfolding other than very
specialized shapes, the suggestion was made in~\cite{do-op02-04} to
seek $k_1 \times k_2 \times k_3$ \emph{refined grid unfoldings},
where every face of the vertex grid is further refined into a grid
of edges.  Positive integers $k_1$, $k_2$ and $k_3$ are associated
with the amount of refinement in the $x$, $y$ and $z$ directions,
respectively; e.g., $z$ perpendicular faces are refined into a $k_1
\times k_2$ grid, and similarly $x$ ($y$) perpendicular faces are
refined into a $k_2 \times k_3$ ($k_1 \times k_3$) grid. It is this
line we pursue in this paper, on a class of shapes not previously
considered.

We define ``Manhattan Tower (MT) polyhedra'' to be the natural
generalization of ``Manhattan Skyline polygons.'' Although we do not
know of an unrefined grid unfolding for this class of shapes, we
prove (Theorem~\ref{theo:5x5}) that there is a $4 \times 5 \times 1$
grid unfolding. Our algorithm peels off a spiral strip that winds
first forward and then interleaves backward around vertical slices
of the polyhedron, recursing as attached slices are encountered.
%

\section{Definitions}
Let $Z_k$ be the plane $\{ z = k \}$, for $k \ge 0$.
Define $\P$ to be a \emph{Manhattan Tower} (MT) if it is an
orthogonal polyhedron such that:
\begin{enumerate}
\squeezelist
\item $\P$ lies in the halfspace $z \ge 0$,
and its intersection with $Z_0$ is a simply
connected orthogonal polygon;
\item For $k < j$, $\P \cap Z_k \supseteq \P \cap Z_{j}$:
the cross-section at higher levels is nested in that for lower levels.
\end{enumerate}

\begin{figure}[htbp]
\centering
\includegraphics[width=0.6\linewidth]{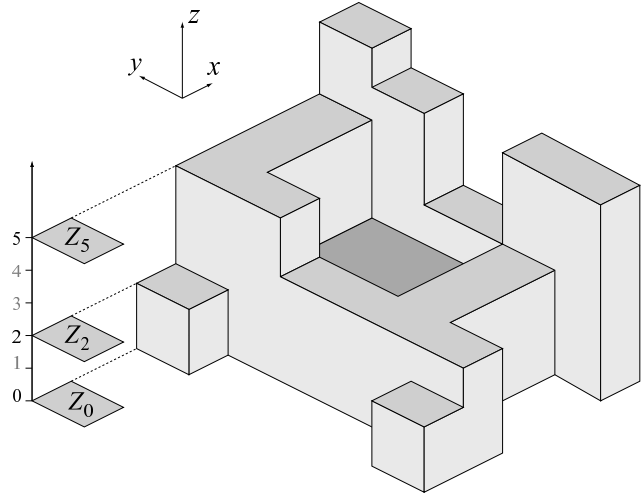}
\caption{Manhattan Tower $\P$.} \label{fig:MTex}
\end{figure}

\begin{figure}[htbp]
\centering
\includegraphics[width=0.90\linewidth]{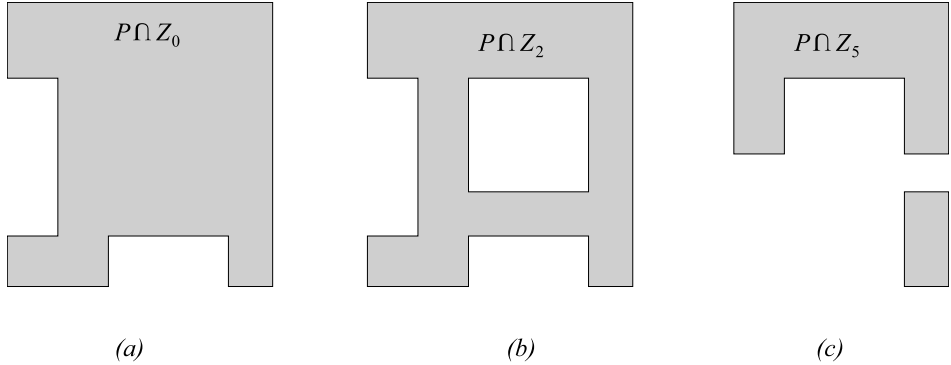}
\caption{ Cross-sections of Manhattan Tower $P$ from
Fig.~\ref{fig:MTex}: (a) $Z_0 \cap P$ is a simple orthogonal
polygon; (b) $Z_2 \cap P$ is an orthogonal polygon with one hole;
(c) $Z_5 \cap P$ has two disjoint components.} \label{fig:Z}
\end{figure}

Manhattan Towers are \emph{terrains} in that they meet each vertical
(parallel to $z$) line in a single segment or not at all; thus they
are \emph{monotone} with respect to $z$.  Fig.~\ref{fig:MTex}a shows
an example. Manhattan Towers may not be monotone with respect to $x$
or $y$, and indeed $\P \cap Z_k$ will in general have several
connected components (cf. Fig.~\ref{fig:Z}c), and may have holes
(cf. Fig.~\ref{fig:Z}b), for $k > 0$.
%
%

As an $xy$-plane sweeps from $Z_0$ upwards, the cross-section of
$\P$ changes at finitely many locations.  Thus a Manhattan Tower
$\P$ may be viewed as consisting of nested layers, with each layer
the extrusion of a set of orthogonal polygons.  The {\em base} of
$\P$ is its bottom layer, which is bounded below by $Z_0$ and above
by the $xy$-plane passing through the first vertex with $z>0$. Note
that, unlike higher layers, the base is simply connected, since it
is an extrusion of $\P \cap Z_0$.

We use the following notation to describe the six types of faces,
depending on the direction in which the outward normal points: {\em
front}: $-y$; {\em back}: $+y$; {\em left}: $-x$; {\em right}: $+x$;
{\em bottom}: $-z$; {\em top}: $+z$. An $x$-$edge$ is an edge that
is parallel to the $x$-axis; $y$-$edges$ and $z$-$edges$ are defined
similarly.

Clockwise (cw) and counterclockwise (ccw) directions are defined
with respect to the viewpoint from $y = -\infty$. Later we will
rotate the coordinate axes in recursive calls, with all terms tied
to the axes altering appropriately.

\section{Recursion Tree}
\begin{figure}[htbp]
\centering
\begin{tabular}{c@{\hspace{0.15\linewidth}}c}
\includegraphics[width=0.55\linewidth]{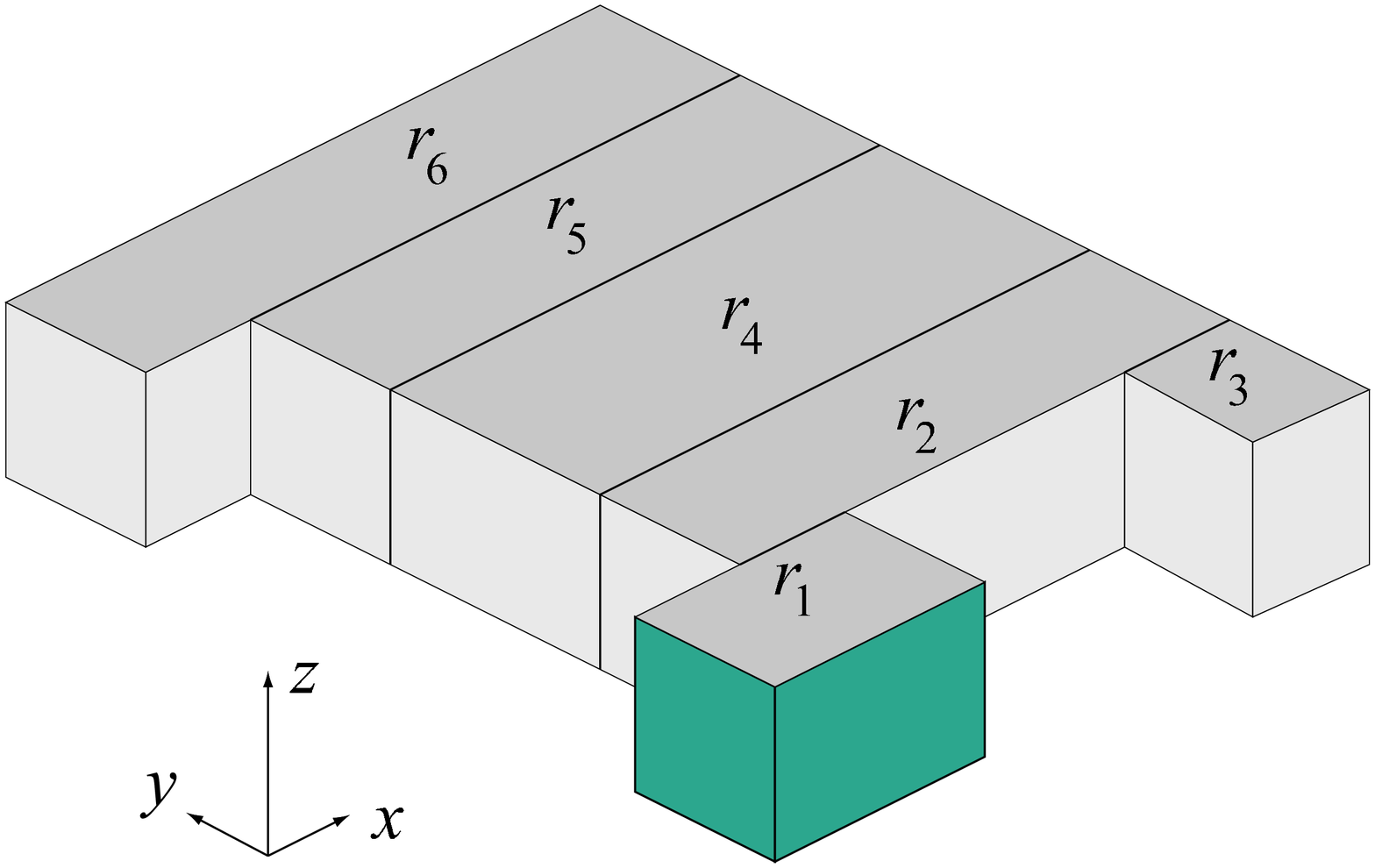} &
\includegraphics[width=0.15\linewidth]{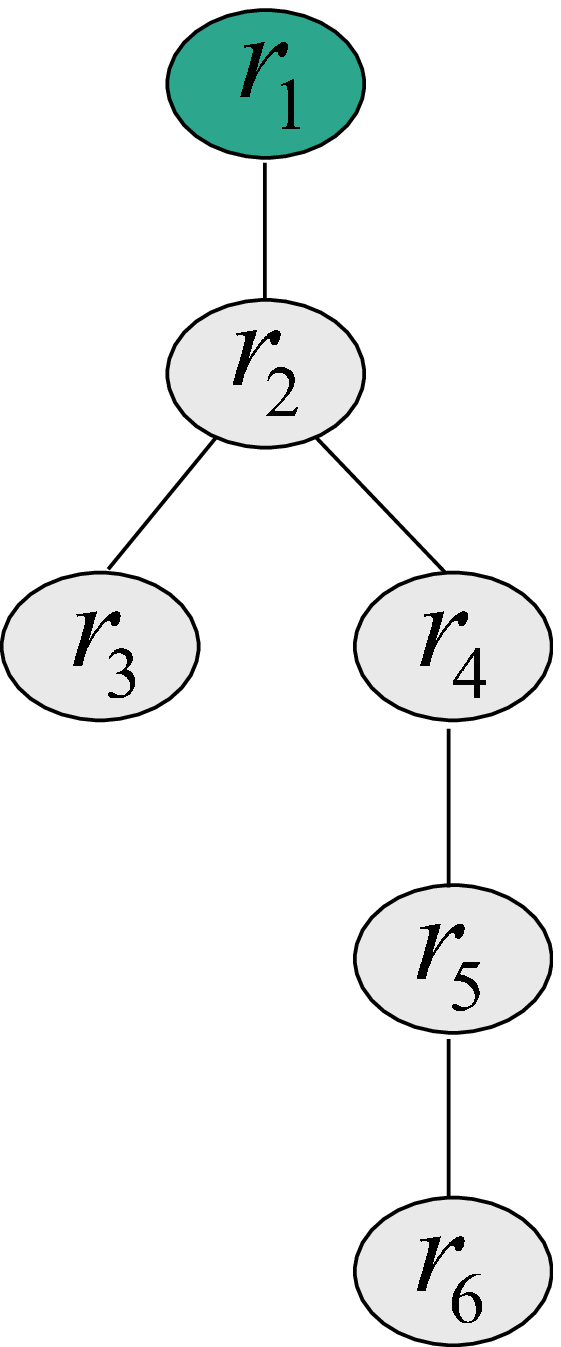}  \\
(a) & (b)
\end{tabular}
\caption{ (a) Partition $\Pi$ of $\P$'s base; (b) Recursion tree
$T$.} \label{fig:basepartition}
\end{figure}

We start with the partition $\Pi$ of the base layer induced by the
$xz$-planes passing through every vertex of $\P$. (The restriction
of the partition to planes orthogonal to $y$ will facilitate
processing in the $\pm y$ directions below.) Such a partition
consists of rectangular boxes only (see
Fig.~\ref{fig:basepartition}a). The dual graph of $\Pi$ has a node
for each box and an edge between each pair of nodes corresponding to
adjacent boxes. Since the base is simply connected, the dual graph
of $\Pi$ is a tree $T$ (Fig.~\ref{fig:basepartition}b), which we
refer to as the {\em recursion tree}. The root of $T$ is a node
corresponding to a box (the \emph{root box}) whose front face has a
minimum $y$-coordinate (with ties arbitrarily broken).
%

It turns out that nearly all unfolding issues are present in
unfolding single-layer MTs, due to the nested-layer structure of
MTs. In Sec.~\ref{sec:TMT} we describe an algorithm for unfolding
single-layer MTs. The algorithm is then extended to handle
multiple-layer MTs in Sec.~\ref{sec:MT}.

\section{$(4 \times 5 \times 1)-$Refined Manhattan Towers}
\begin{figure}[htbp]
\centering
\begin{tabular}{c@{\hspace{0.05\linewidth}}c}
\includegraphics[width=0.5\linewidth]{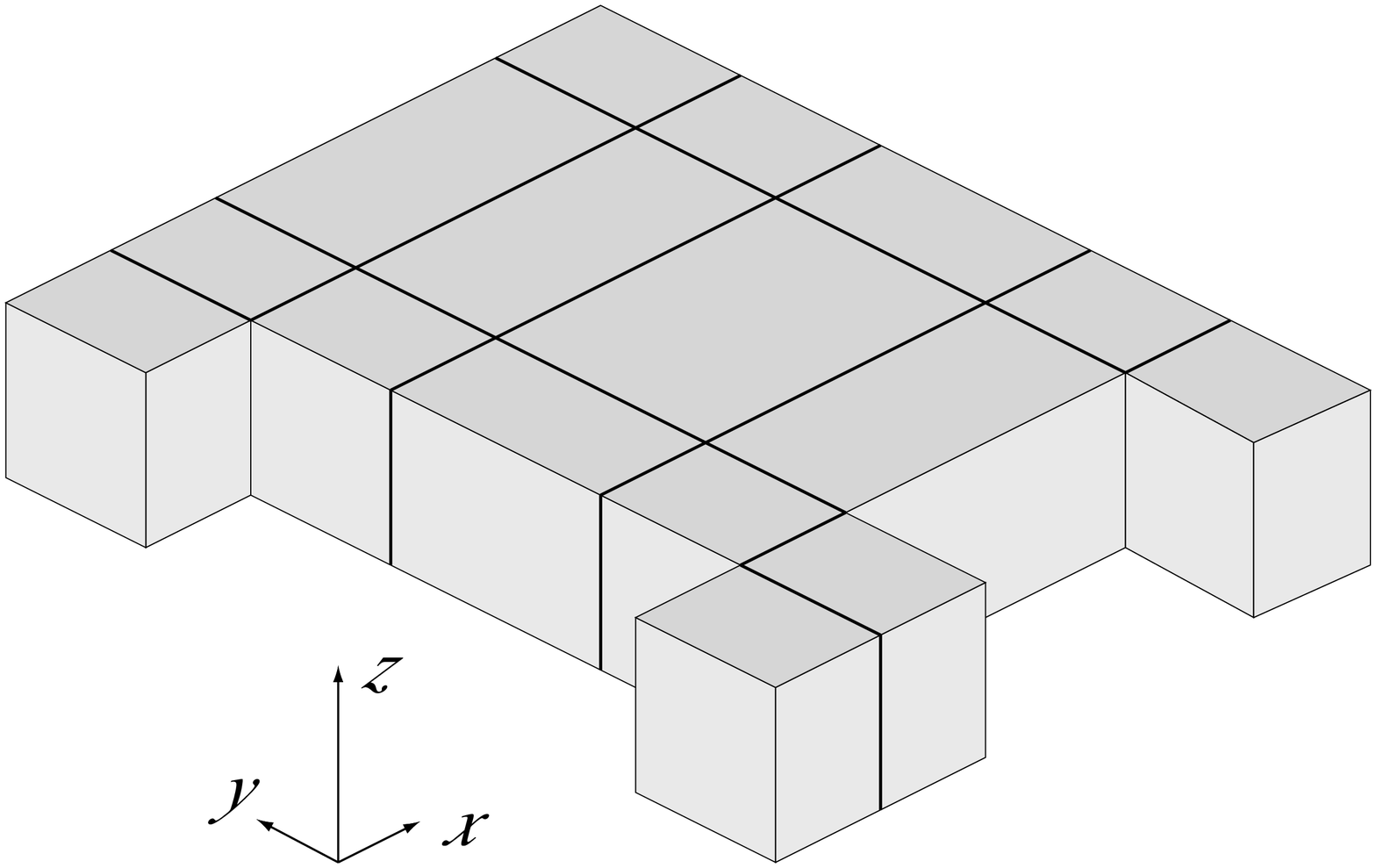} &
\includegraphics[width=0.5\linewidth]{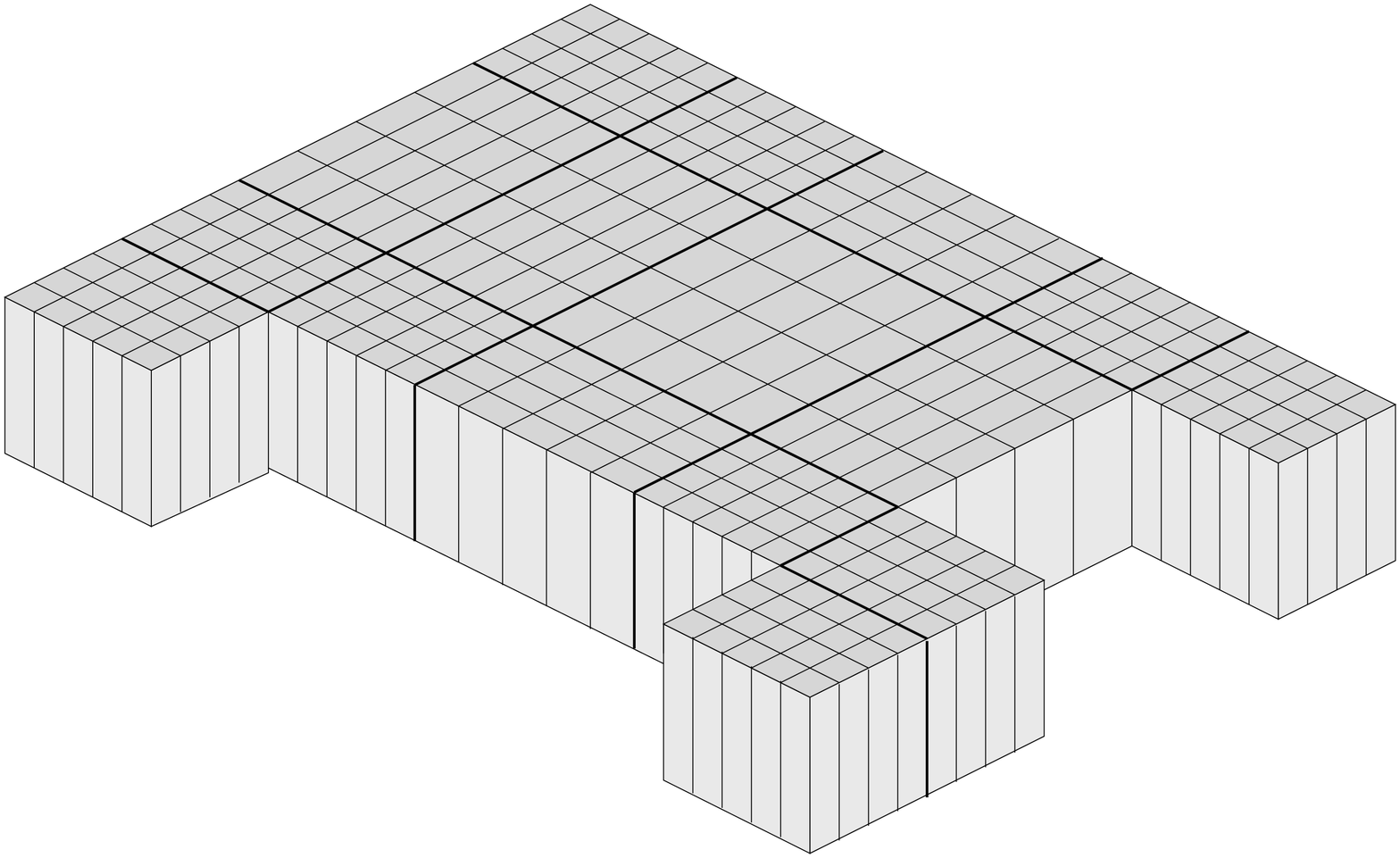} \\
(a) & (b)
\end{tabular}
\caption{ (a) Gridded MT base; (b) $(4 \times 5 \times 1)$-refined MT
base.} \label{fig:basepartition2}
\end{figure}

Fig.~\ref{fig:basepartition2} illustrates the refinement process,
using the base from Fig.~\ref{fig:basepartition}a as an example.
%
%
The \emph{gridded} base (Fig.~\ref{fig:basepartition2}a) contains
additional surface edges induced by $yz$-coordinate planes through
each vertex. A $4 \times 5 \times 1$ refinement of the gridded base
further partitions each horizontal grid rectangle into a $4 \times
5$ grid. In addition to gridedges of the gridded base, the $(4
\times 5 \times 1)$-\emph{refined} base
(Fig.~\ref{fig:basepartition2}b) contains all surface edges induced
by coordinate planes passing through each gridpoint in the
refinement.
In the following we show that every $(4 \times 5 \times
1)$-refined MT can be edge-unfolded.

\section{Single-Layer MTs}
\label{sec:TMT} A single-layer Manhattan Tower consists of a single
layer, the {\em base} layer. We describe the unfolding algorithm
recursively, starting with the base case in which the layer is a
single rectangular box.

\subsection{Single Box Unfolding}
\label{sec:basecase}
\begin{figure}[htbp]
\centering
\includegraphics[width=0.95\linewidth]{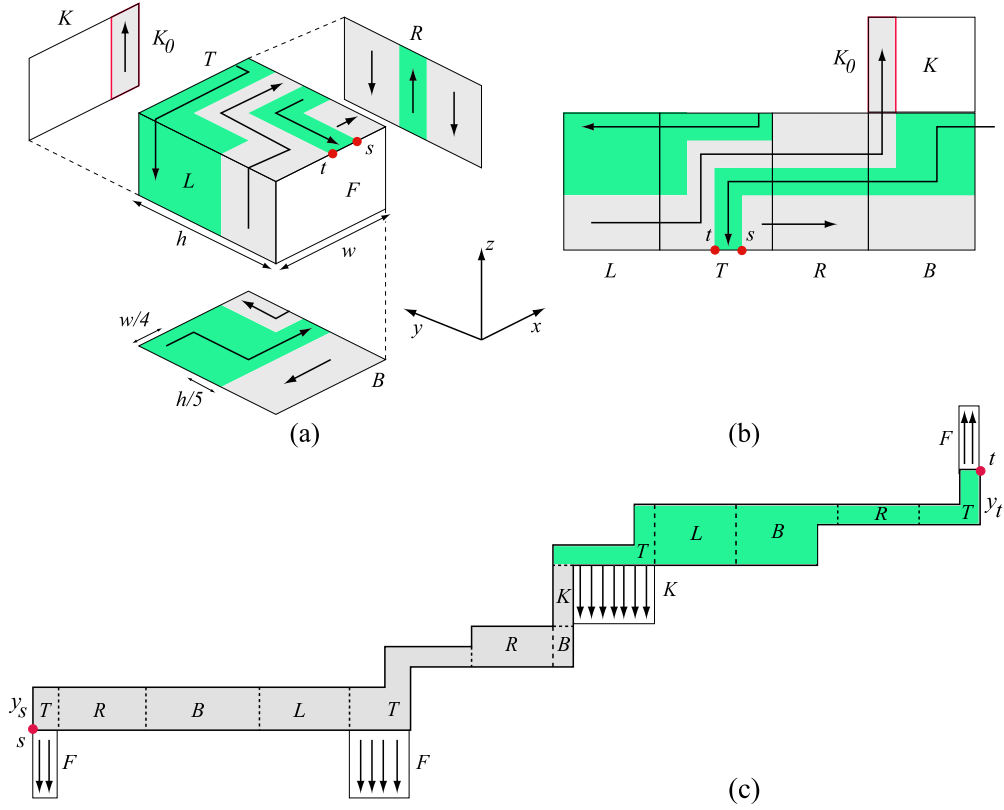}
\caption{Single box unfolding: $s$ adjacent to $t$ (a) Front view of
box $r$ and mirror view of right ($R$), bottom ($B$) and back ($K$)
faces, marked with unfolding cuts (b) Faces of $r$ flattened out
(front face not shown) (c) Spiral unfolding of $r$; labels identify
faces containing the unfolded pieces.} \label{fig:basecase.a}
\end{figure}

\begin{figure}[htbp]
\centering
\includegraphics[width=0.95\linewidth]{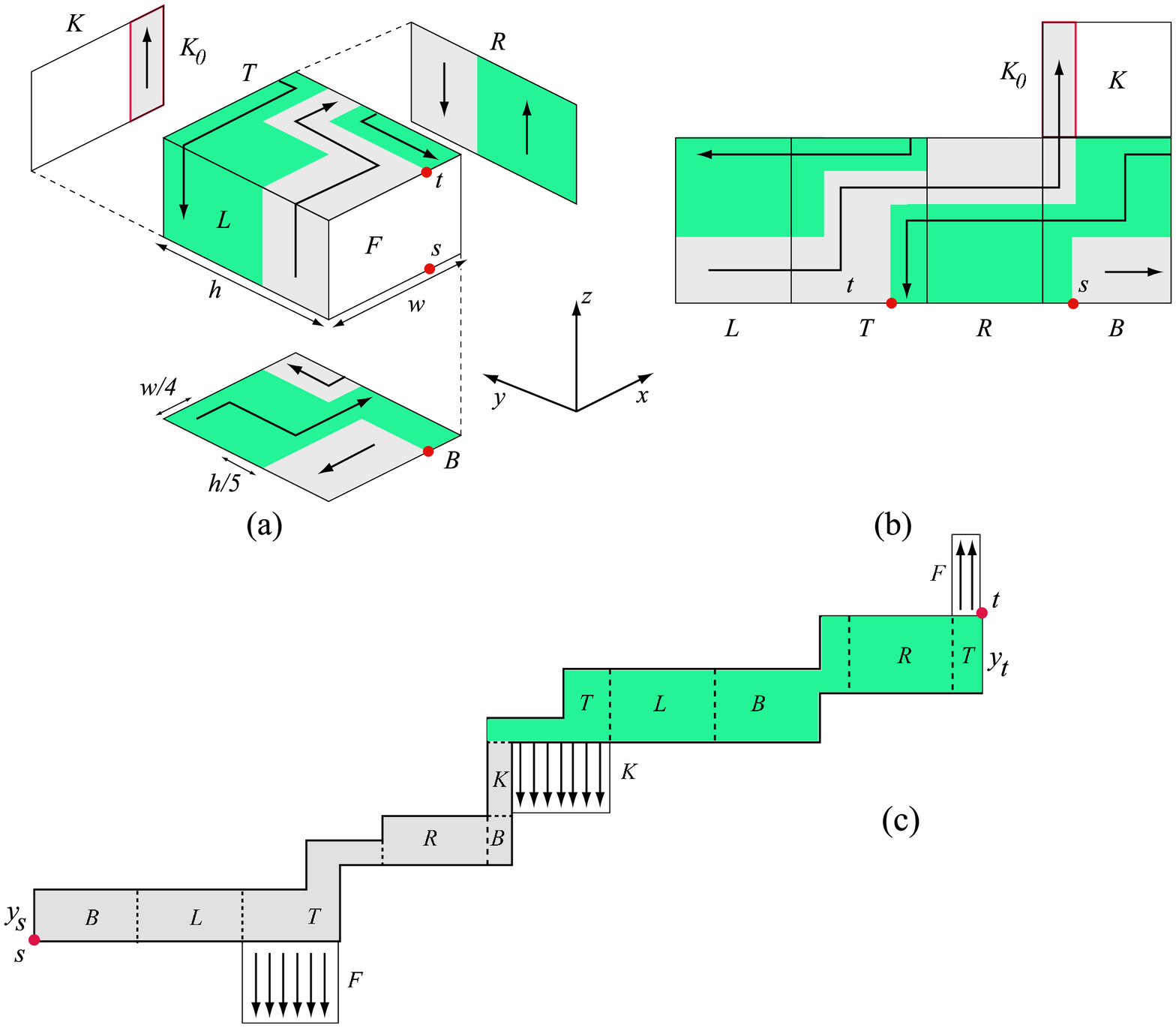}
\caption{Single box unfolding: $s$ and $t$ on opposite front edges
(a) Front view of box $r$ and mirror view of right ($R$), bottom
($B$) and back ($K$) faces, marked with unfolding cuts (b) Faces of
$r$ flattened out (front face not shown) (c) Spiral unfolding of
$r$; labels identify faces containing the unfolded pieces.}
\label{fig:basecase.b}
\end{figure}

Let $r$ be a $(4 \times 5 \times 1)$-gridded rectangular box and let
$T$, $R$, $B$, $L$, $K$ and $F$ be the top, right, bottom, left,
back and front faces of $r$, respectively. Let $s$ and $t$ be two
gridpoints either adjacent on the same $x$-edge of $r$ (as in
Fig.~\ref{fig:basecase.a}a), or one on the top $x$-edge and one on
the bottom $x$-edge of the front face of $r$ (as in
Fig~\ref{fig:basecase.b}a). Let $y_s$ and $y_t$ be the ($y$ parallel)
gridedges
incident to $s$ and $t$. The unfolding of $r$ starts at $y_s$ and
ends at $y_t$. More precisely, this means the following. Let
$\xi_{2d}$ ($\xi_{3d}$) denote the planar (three-dimensional)
embedding of the cut surface piece.
Then
$\xi_{2d}$ has $y_s$ on its far left and $y_t$ on its far right (as
in Figs.~\ref{fig:basecase.a}c and~\ref{fig:basecase.b}c).

%

The main unfolding idea  is to cut the top, right, bottom and left
faces so that they unfold into a staircase-like strip and attach front and back
faces to it vertically without overlap. We collectively refer to
the top, right, bottom and left faces as \emph{support} faces
(intuitively, they support the front and back faces). Roughly
stated, $\xi_{3d}$ starts at $y_s$, spirals cw around the support
faces toward the back face, crosses the back face, then spirals ccw
around the support faces back to $y_t$. This idea is illustrated in
Figs.~\ref{fig:basecase.a} and~\ref{fig:basecase.b}. In the
following we provide the details for the case when $s$ and $t$ are
adjacent on the top front edge of $r$ (Fig.~\ref{fig:basecase.a}).
The case when $s$ lies on a bottom front edge and $t$ lies on a top
front edge of $r$ is similar and is illustrated in
Fig.~\ref{fig:basecase.b}; the case when $s$ is on the top and $t$
is on the bottom is identical, when viewed through an $xy$-mirror.

As illustrated in Fig. 5, let $w$ be the $x$-extent and let $h$ be the
$y$-extent of $r$. We implicitly define the unfolding cuts by
describing the surface pieces encountered in a walk along $\xi_{3d}$
on the surface of $r$ (delineated by unfolding cuts). Starting at
$y_s$, walk cw along a rectangular strip of $y$-extent equal to $2h/5$
(two gridfaces-wide) that spirals around the support faces from $y_s$
to $y_t$. This spiral strip lies adjacent to the front face of $r$; we
will refer to it as the \emph{front spiral} of $\xi_{3d}$. At $y_t$,
take a left turn and continue along a rectangular strip (orthogonal to
the front spiral and right-aligned at $t$) of $y$-extent equal to
$2h/5$ (two gridfaces-wide) and $x$-extent equal to $w/4$ (one
gridface-long). At the end of this strip, take a right turn and
continue along a rectangular strip of $y$-extent equal to $h/5$, until
the right face $R$ is met; at this point, the strip thickens to a
$y$-extent equal to $2h/5$ (two gridfaces-wide), so that it touches the
back face $K$ of the box. The strip touching $K$ consumes the entire
length of the right face $R$, plus an additional $w/4$ (one gridface)
amount onto the adjacent bottom face $B$. At the end of this bottom
strip, take a left turn and continue along a $w/4$-wide strip across
back face $K$ and up onto the top face $T$.  The piece of $\xi_{3d}$
traversed so far is called the \emph{forward} spiral; the remaining
piece is called the \emph{backward} spiral, conveying the fact that
from this point on $\xi_{3d}$ spirals ccw around the support faces
back to $y_t$. The piece of the backward spiral adjacent to the back
face is the \emph{back spiral} of $r$. The planar piece $\xi_{2d}$
(obtained by laying $\xi_{3d}$ out in the plane) has the
staircase-like shape illustrated in
Fig~\ref{fig:basecase.a}c. Conceptually, the front face $F$ and the
back face $K$ are not part of the unfolding described so far; however,
they can be flipped up and attached vertically to $\xi_{2d}$ without
overlap (see the striped faces in Fig.~\ref{fig:basecase.a}c), a point
to which we return in Sec.~\ref{sec:front.back}.

\subsection{Recursion Structure}
\label{sec:Recursion}
In general, a box $r$ has children (adjacent boxes) attached along
its front and/or back face. Call a child attached on the front a
{\em front child} and a child attached on the back a {\em back
child}. In unfolding $r$, we unwind the support (top, bottom, left,
right) faces into a staircase-like strip just as described for the
single box. But when the front/back spiral runs alongside the
front/back face of $r$ and encounters an adjacent child, the
unfolding of $r$ is temporarily suspended, the child is recursively
unfolded, then the unfolding of $r$ resumes where it left off.

At any time in the recursive algorithm there is a \emph{forward}
direction, corresponding to the initial spiraling from front to back
(the lighter strip in Figs.~\ref{fig:basecase.a}
and~\ref{fig:basecase.b}), and an opposing \emph{backward} direction
corresponding to the subsequent reverse spiraling from back to front
(the darker strip in Figs.~\ref{fig:basecase.a}
and~\ref{fig:basecase.b}). When the recursion processes a front
child, the sense of forward/backward is reversed: we view the
coordinate system rotated so that the $+y$ axis is aligned with the
forward direction of the child's spiral, with all terms tied to the
axes altering appropriately. In particular, this means that the
start and end unfolding points $s'$, $t'$ of a front child $r'$ lie
on the front face of $r'$, as defined in the rotated system.

\begin{figure}[htbp]
\centering
\includegraphics[width=0.6\linewidth]{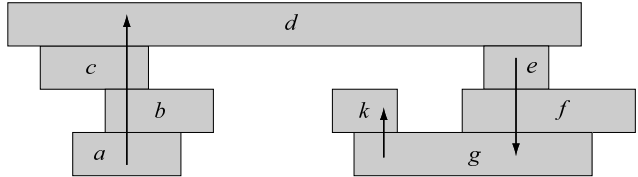}
\caption{Arrows indicate which direction is \emph{forward} in the
recursive processing.}
\label{fig:forward.recursion}
\end{figure}

For example, in Fig.~\ref{fig:forward.recursion}, boxes $a$, $b$,
$c$, $d$ are processed from front to back. But recursion on $e$, a
front child of $d$, reverses the sense of forward, which continues
through $e$, $f$, and $g$. We can view the coordinate system rotated
so that $+y$ is aligned with the arrows shown.  Thus $f$ is a back
child of $e$, $g$ is a back child of $f$, and $k$ a front child of
$g$.  Again the sense of forward is reversed for the processing of
$k$.

\subsection{Suturing Techniques}
\label{sec:suturing}

We employ two methods to ``suture'' a child's unfolding to its
parent's unfolding. The first method,
{\em same-direction suture}, is used to suture all front children to
their parent. If there are no back children, then a strip from the
back face of the parent ($K_0$ in Figs~\ref{fig:basecase.a}
and~\ref{fig:basecase.b}) is used to reverse the direction of the
spiral to complete the parent's unfolding, as described in
Sec.~\ref{sec:basecase} for the single box. However, if the
parent has one or more back children, these children cover parts or
perhaps all of the back face of the parent, and the back face strip
may not be available for the reversal. So instead we use a second
suturing method, {\em reverse-direction suture}, for one of the back
children. This suture uses the child's unfolding to reverse the
direction of the parent's spiral, and does not require a back-face
strip. We choose {\em exactly one} back child for reverse-direction
suturing.  Although any such child would serve, for definiteness we
select the rightmost child. Our suturing rules are as follows:
\begin{enumerate}
\squeezelist
\item For every front child, use same-direction suturing.
\item For the rightmost back child, use reverse-direction suturing.
\item For remaining back children, use same-direction suturing.
\end{enumerate}

\vspace{-0.15in}
\subsubsection{Same-direction suture}
\label{sec:same.direction}
We first note that a front child $r'$ never entirely covers the
front face of its parent box $r$, because the parent of $r$ is also
adjacent to the front face of $r$. This is evident in
Fig.~\ref{fig:forward.recursion}, where $e$ cannot cover the front
face of $d$ because $d$'s parent, $c$, is also adjacent along that
side. Similarly, $k$ cannot cover the ``front'' face of $g$ (where
here the sense of front is reversed with the forward direction of
processing) because $g$'s parent $f$ is also adjacent along that
side. The same-direction suture may only be applied in such a
situation of non-complete coverage of the shared front face, for it
uses a thin (one gridface-wide) vertical strip off that face.
%
\begin{figure}[htbp]
  \centering
  \includegraphics[width=0.74\linewidth]{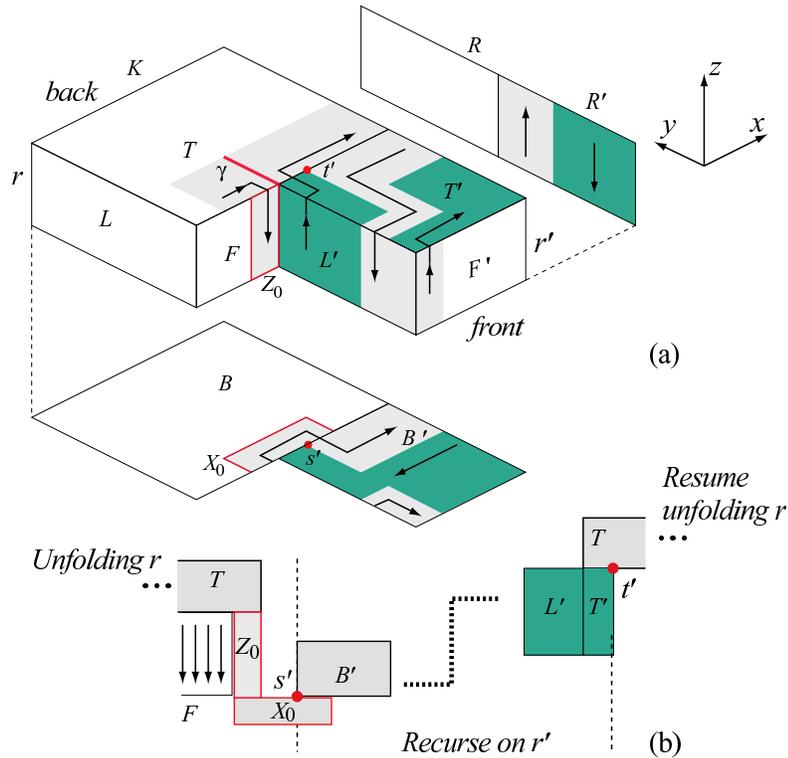}
  \caption{Same-direction suture.
           (a) Front view of faces root box $r$ and front child $r'$,
           with mirror bottom, left and back views.
           (b) Result $\xi_{2d}$ of recursive unfolding.}
  \label{fig:same.suture}
\end{figure}

This suture begins at the point where the parent's spiral meets an
adjacent child as it runs alongside its front or back face. To be
more specific, consider the case when $r'$ is a front child of $r$,
and the parent's front spiral meets $r'$ as it runs along the top of
$r$. This situation is illustrated in Fig.~\ref{fig:same.suture}.
The same-direction suture begins by cutting a vertical strip $I$ off
the front face of parent $r$, which includes all vertical gridfaces
that lie alongside child $r'$ (see Fig.~\ref{fig:same.suture}a),
then it takes a bite $J$ one gridface-thick and three gridfaces-long
(in the $x$-direction) off the bottom face of the parent. This marks
the gridedge $y_{s'}$ on $r'$ where the child's spiral unfolding
starts.
The child's spiral unfolding ends at top gridedge $y_{t'}$ of the
same $x$-coordinate as $y_{s'}$. When the child's unfolding is
complete, the spiral unfolding of the parent resumes at the
$y$-gridedge it left off (see the cut labeled $\gamma$ in
Fig.~\ref{fig:same.suture}). The other cases are similar: if $r'$ is
a back child of $r$, $I$ occurs on the back face of $r$; and if the
parent's front spiral meets $r'$ as it runs along the bottom of
$r$,\footnote{This only happens if $r'$ is a front child of $r$.}
$J$ occurs on the top face of $r$ (see child $r_4$ and parent $r_2$
in Fig.~\ref{fig:3hshape}). It is this last case that requires a $5$
refinement in the $y$ direction: the front spiral must be two
gridfaces-thick so that cutting $J$ off it does not disconnect it.

As the name suggests, this suturing technique preserves the
unwinding direction (cw or ccw) of the parent's spiral.
In Fig.~\ref{fig:same.suture}, notice that the parent's spiral
unfolds in cw direction on top face $T$ before the suture begins.
The parent's cw unfolding is suspended at $y$-gridedge marked
$\gamma$, and after the child is unfolded, the parent's spiral
resumes its unfolding in cw direction at $\gamma$. The unfolded
surface $\xi_{2d}$ is shown in
Fig.~\ref{fig:same.suture}b.

\subsubsection{Reverse-direction suture}
\label{sec:rev.direction}

\begin{figure}[htbp]
  \centering
  \includegraphics[width=0.86\linewidth]{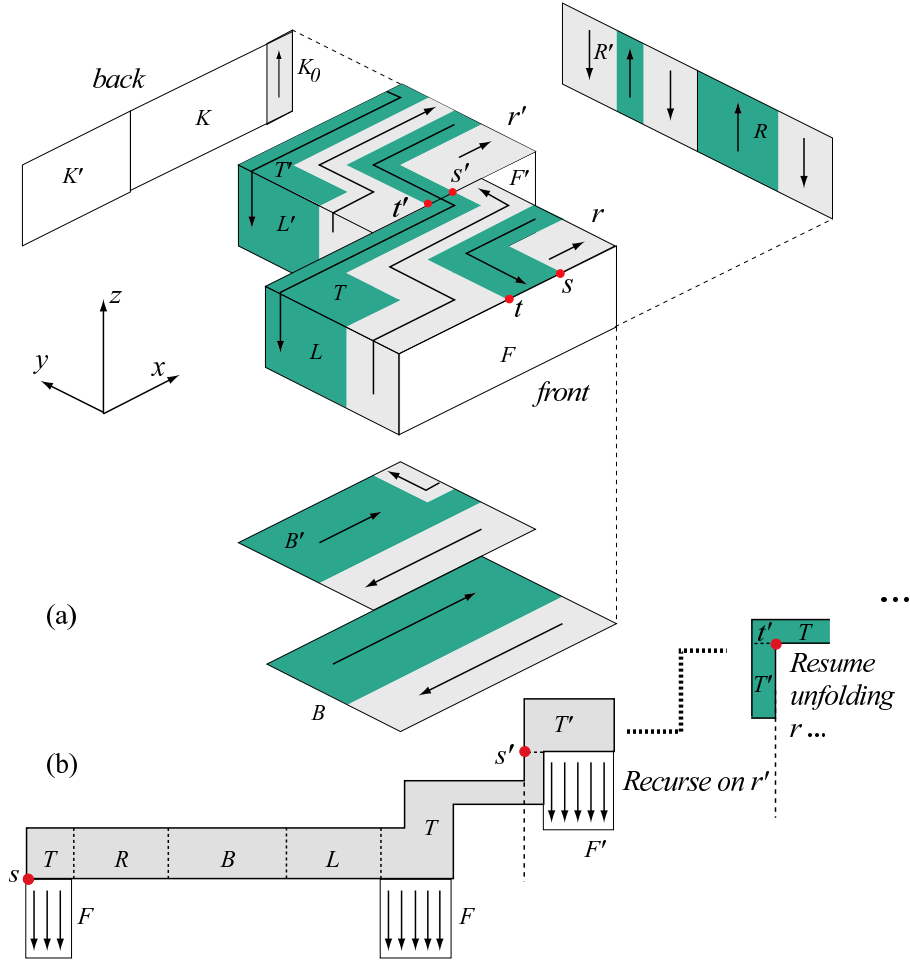}
  \caption{Reverse-direction suture. (a)
           Front view of faces of root box $r$ and back child $r'$,
           with mirror bottom, left and back views.
           (b) Result $\xi_{2d}$ of recursive unfolding.}
  \label{fig:rev.suture}
\end{figure}

This suture begins after the parent's spiral completes its first
cycle around the support (top, right, bottom, left) faces, as
illustrated in Fig.~\ref{fig:rev.suture} for parent $r$ and back
child $r'$. As in the single box case (Sec.~\ref{sec:basecase}),
after a forward move in the $+y$-direction, the spiral starts a
second cycle around the support faces. However, unlike in the single
box case, the spiral stops as soon as it reaches a $y$-gridedge of
the same $x$-coordinate as the rightmost gridpoint $u$ that the
parent shares with a back child. At that point, the parent's spiral
continues with a gridface-thick strip $S$ in the $+y$-direction,
right-aligned at $y_u$. Let $s'$ be the left corner of $S$ on the
boundary shared by $r$ and $r'$. The unfolding of $r'$ begins at
gridedge $y_{s'}$ and ends at gridedge $y_{t'}$ immediately to the
left of $y_{s'}$ on top of $r'$. When the child's unfolding is
complete, the unfolding of the parent resumes at the gridedge it
left off, with the spiral unwinding in reverse direction.
%

As the name suggests, this suturing technique reverses the unwinding
direction (cw or ccw) of the parent's spiral. In
Fig.~\ref{fig:rev.suture}, notice that the parent's spiral unfolds
in cw direction on top face $T$ before the suture begins. After the
child is unfolded, the parent's spiral resumes its unfolding in ccw
direction at $y_{s'}$. The result $\xi_{2d}$ of this unfolding is
shown in Fig.~\ref{fig:rev.suture}b.

\subsection{Attaching Front and Back Faces}
\label{sec:front.back} The spiral strip $\xi_{3d}$ covers all of the
top, bottom, right, and left faces of the base. It also covers the
gridface-thick strips of a front/back face used by the
same-direction sutures ($I$ in Fig.~\ref{fig:same.suture}) and the
gridface-thick strips of back faces used to reverse the spiral
direction in the base cases ($K_0$ in Figs~\ref{fig:basecase.a}
and~\ref{fig:basecase.b}). The staircase structure of $\xi_{2d}$
(shown formally in Theorem~\ref{thm:correctedness}) guarantees that
no overlap occurs.

We now show that remaining exposed front and back pieces that are
not part of $\xi_{3d}$ can be attached orthogonally to $\xi_{2d}$
without overlap. Consider the set of top gridedges shared by top
faces with front/back faces. These gridedges occur on the horizontal
boundaries of $\xi_{2d}$ as a collection of one or more contiguous
segments. We partition the front/back faces by imagining these top
gridedges illuminate downward lightrays on front/back faces. Then
all front and back pieces are illuminated, and these pieces are
attached to corresponding illuminating gridedges (see
Figs.~\ref{fig:basecase.a}c,~\ref{fig:basecase.b}c,~\ref{fig:same.suture}b
and~\ref{fig:rev.suture}b). Although no interior points overlap in
the unfolding, we allow {\em edge overlap}, which corresponds to the
physical model of cutting out the unfolded piece from a sheet of
paper. For example, in Fig.~\ref{fig:rev.suture}b a left gridedge of
$F'$ overlaps a gridedge of $\xi_{2d}$. It is not difficult to avoid
edge overlap (e.g. by making the portion of the strip causing the
edge overlap narrower to separate it from $F'$), but doing so
requires increasing the degree of refinement.

The next section summarizes the entire unfolding process for
single-layer MTs.

\subsection{Unfolding Algorithm for Single-Layer MTs}
\label{sec:base.algorithm}

Consider an arbitrary base partitioned into rectangular boxes with
$xz$-planes $Y_0, Y_1, \ldots$ through each vertex. Select a root
box $r$ adjacent to $Y_0$ (breaking ties arbitrarily) and set the
forward unwinding direction $d$ to be cw. Let $y_s$ and $y_t$ be top
$y$-gridedges of $r$, as described in Sec.~\ref{sec:basecase} for
the single-box case. Our recursive unfolding starts at root box $r$
and proceeds as follows.

\begin{center}
\vbox{\hrule width\linewidth} \noindent {\tt Algorithm UNFOLD($r,
y_s, y_t$)} \vbox{\hrule width\linewidth} {\small
\begin{tabbing}
.....\=.......\=.........\=.......\=.......\=
........................................................................\=...\kill
1. \> Start unfolding the forward spiral piece adjacent to front
face (\S~\ref{sec:basecase}). \\
\\
2. \> {\bf Unfolding Front Children.}
    For each front child $r'$ of $r$ encountered \\
    \>\> Determine gridedges $y_{s'}$, $y_{t'}$ using same-direction suture (\S~\ref{sec:same.direction}). \\
    \>\> Recurse: UNFOLD($r', y_{s'}, y_{t'}$). \\
\\
3. \> If $r$ has no back children then complete the unfolding of $r$ (~\S~\ref{sec:basecase}) and exit. \\
4. \> Determine start and end gridedges $y_{s'}$, $y_{t'}$ for
rightmost back child $r'$ \\
\> using reverse-direction suture (\S~\ref{sec:rev.direction}). \\
5. \> Complete the unfolding of the forward spiral up to $y_{s'}$
(\S~\ref{sec:rev.direction}). \\
6. \> Recurse: UNFOLD($r', y_{s'}, y_{t'}$). \\
7. \> Continue unfolding the back spiral adjacent to back face (\S~\ref{sec:basecase}). \\
\\
8. \> {\bf Unfolding Rest of Back Children.}
    For each back child $r'$ of $r$ encountered \\
    \>\> Determine gridedges $y_{s'}$, $y_{t'}$ using same-direction suture (\S~\ref{sec:same.direction}). \\
    \>\> Recurse: UNFOLD($r', y_{s'}, y_{t'}$). \\
    \\
9. \> Complete the unfolding of $r$ by spiraling back to $y_t$ (\S~\ref{sec:basecase}). \\
10. \> Hang front and back faces off the unfolded spiral.
(\S~\ref{sec:front.back}).
\end{tabbing}}
\vbox{\hrule width\linewidth}
\end{center}

\begin{figure}[htbp]
\centering
\includegraphics[width=\linewidth]{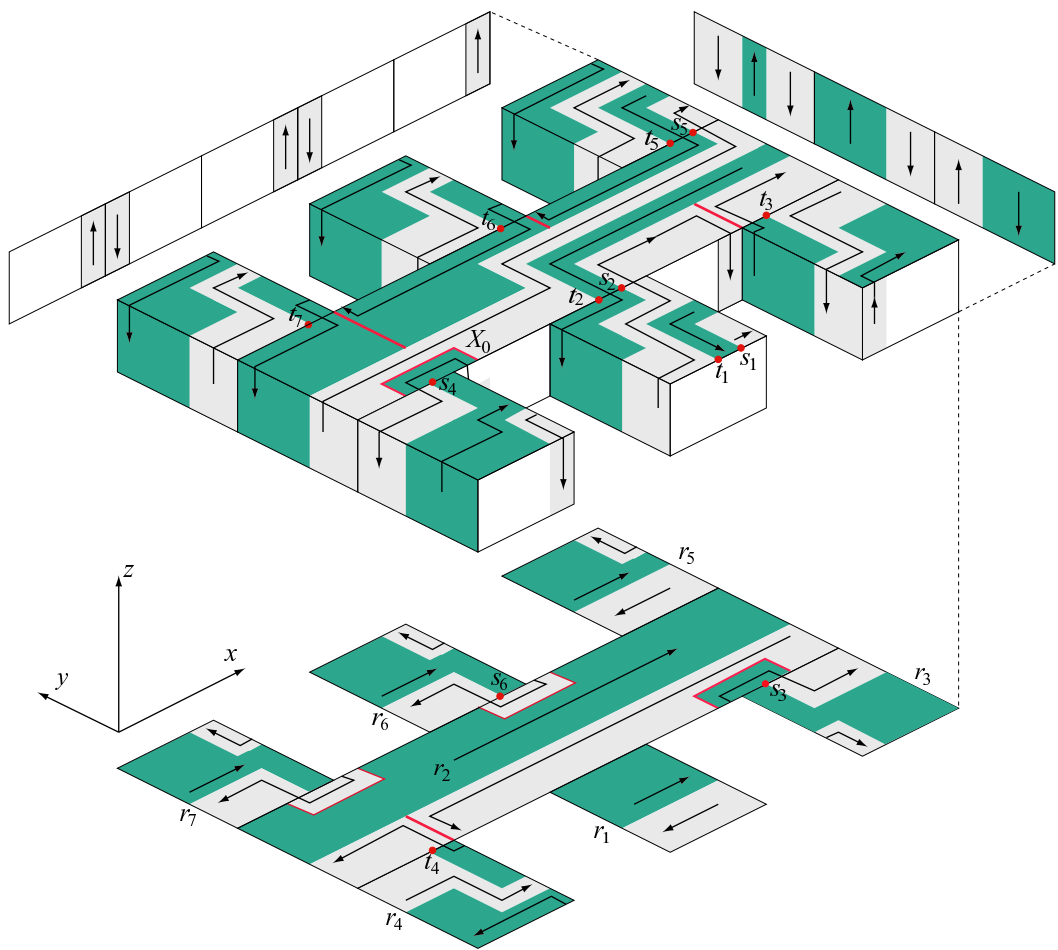}
\caption{Unfolding a 3-legged $H$-shaped base.} \label{fig:3hshape}
\end{figure}

This algorithm can be easily implemented to run in $O(n^2)$ time on
a polyhedron $\P$ with $n$ vertices.
Fig.~\ref{fig:3hshape} illustrates the recursive unfolding algorithm
on a $3$-legged $H$-shaped base. The unfolding starts at gridedge
$y_{s_1}$ of root box $r_1$ and ends at gridedge $y_{t_1}$.
(Only the endpoints $s_1$ and $t_1$ of these two
gridedges are marked in Fig~\ref{fig:3hshape}.) The
spiral strip encounters the boxes in the order $r_1, r_2, r_3, r_4,
r_5, r_6$ and $r_7$, which corresponds to the ordering of the
recursive calls. For each $i$, $y_{s_i}$ and $y_{t_i}$ are gridedges
of $r_i$ where the unfolding of $r_i$ starts and ends.The algorithm
uses reverse-direction suture to attach back child $r_2$ to parent
$r_1$; same-direction suture to attach front child $r_3$, and then
$r_4$, to parent $r_2$;
reverse-direction suture
to attach back child $r_5$ to parent $r_2$;
and same-direction suture to
attach back child $r_6$, and then $r_7$, to parent $r_2$.
Note that a refinement of 5 in the $y$ direction
is necessary on top of box $r_2$ for this unfolding.
%

\begin{thm}
The {\tt UNFOLD}($r,y_s,y_t$) algorithm unfolds all boxes in the
recursion tree rooted at $r$ into a staircase-like strip $\xi_{2d}$
completely contained between the vertical lines passing through
$y_s$ and $y_t$. \label{thm:correctedness}
\end{thm}
\begin{pf}
The proof is by induction on the height $k$ of the recursion tree
rooted at $r$. The base case is $k = 0$ and corresponds to single
node trees. This is the case illustrated in
Figs.~\ref{fig:basecase.a} and~\ref{fig:basecase.b}, which satisfy
the claim of the theorem.

The inductive hypothesis is that the theorem is true for any
recursion tree of height $k-1$ or less. To prove the inductive step,
consider a recursion tree $T$ of height $k$ rooted at $r$. The
staircase strip $\xi_{2d}(r)$ of $r$ alone, ignoring all children,
fits between the vertical lines passing through $y_s$ and $y_t$ (cf.
Figs.~\ref{fig:basecase.a}c and~\ref{fig:basecase.b}c).

Assume, w.l.o.g., that $r$ unfolds cw. There are two possible
placements of $s$ and $t$ on $r$: (i) $s$ and $t$ are on {\em
opposite} top/bottom edges of the front face of $r$
(Fig.~\ref{fig:basecase.b}a), as placed by a same-direction suture,
or (ii) $s$ and $t$ are on a {\em same} top/bottom edge of $r$
(Fig.~\ref{fig:basecase.a}a), as placed by a reverse-direction
suture. In either case, $s$ and $t$ are placed in such a way that no
children exist along the path extending cw from $t$ to $s$ on $r$.
This means that all front children of $r$ are encountered during the
unwinding of $r$'s front spiral from $s$ to $t$ on $r$. That all
back children are encountered during the unwinding of $r$'s back
spiral is clear: starting at the rightmost back child, the back
spiral makes a complete cycle around the back face.

Consider now an arbitrary child $r'$ of $r$ in $T$ and let $T'$ be
the subtree rooted at $r'$. As noted above, $r'$ will be encountered
during the unfolding of $r$. Let $y_{s'}$ and $y_{t'}$ be the
gridedges on $r'$ where the unfolding of $r'$ starts and ends. The
inductive hypothesis applied on $T'$ tells us that the strip
$\xi_{2d}(r')$ corresponding to $T'$ fits between the vertical lines
passing through $y_{s'}$ and $y_{t'}$.
Fig.~\ref{fig:same.suture}b illustrates the same-direction suture:
when $\xi_{2d}(r')$ is sutured to $\xi_{2d}(r)$, the strip
$\xi_{2d}(r)$ expands horizontally and remains contained between the
vertical lines passing through $y_s$ and $y_t$. The
reverse-unfolding suture has a similar behavior (illustrated in
Fig.~\ref{fig:rev.suture}b), thus completing this proof.
\end{pf}

\section{Multiple-Layer MTs}
\label{sec:MT} 
\begin{figure}[htbp]
\centering
\includegraphics[width=0.78\linewidth]{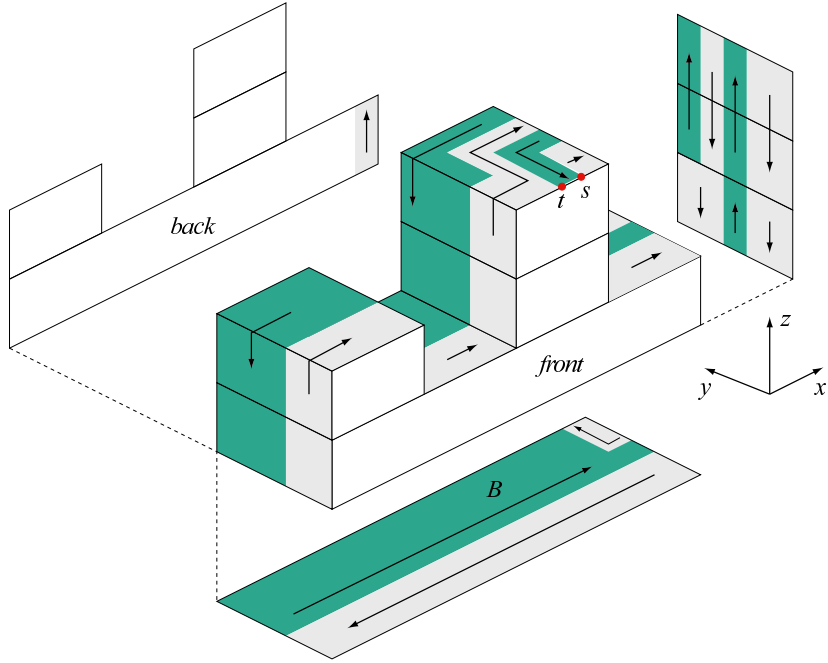}
\caption{Front view of single slab $S(r)$, with mirror bottom, left
and back views.} \label{fig:basecase2}
\end{figure}

Few changes are necessary to make the single-layer
unfolding algorithm from Sec.~\ref{sec:base.algorithm} handle
multiple-layer Manhattan Towers. In fact, the view of the cuts used
to form $\xi_{3d}$ from $z = \pm \infty$ in the multi-layer case is
identical to that in the single-layer unfolding. All the differences
lie in vertical ($z$-parallel) strips used to adjust for differing
tower heights. When there are multiple-layers, the basic unit to
unfold is a vertical {\em slab} $S(r)$ consisting of a box $r$ in
the partition $\Pi$ of the base layer and all the towers that rest
on top of $r$ (see Fig.~\ref{fig:basecase2}).
%
A slab is a Manhattan Skyline polygon parallel to the $xz$-plane
extruded in the $y$ direction: the projection of the top faces of
the slab on the $xy$-plane forms a partition of the (unique) bottom
face (face $B$ in Fig.~\ref{fig:basecase2}). It is here that we make
essential use of the assumptions that $\P \cap Z_0$ is a simply
connected orthogonal polygon, and the cross-sections at higher
levels are nested in those for lower levels.

The unfolding of a slab $S(r)$ is similar to the unfolding of a
single box:
\begin{enumerate}
\item Select an arbitrary top face $T$ of the slab.
\item Select start and end gridedges $y_s$ and $y_t$ on $T$ as in the single box
case.
\item Unfold $S(r)$ using the procedure described in
Sec.~\ref{sec:basecase} for $r$.
\end{enumerate}
The only difference is that a slab may have multiple left/right/top
faces, causing the spiral $\xi_{3d}$ to cycle up and down over the
towers of $S(r)$, as illustrated in Fig.~\ref{fig:basecase2}. As a
result, $\xi_{2d}$ lengthens horizontally, but still maintaining its
staircase structure. As in the case of a single box, $\xi_{3d}$
covers all of the top, right, bottom and left faces. The remaining
front and back pieces are attached to $\xi_{2d}$ using the
illumination scheme described in Sec.~\ref{sec:front.back}.
%
\begin{figure}[htbp]
\centering
\includegraphics[width=0.98\linewidth]{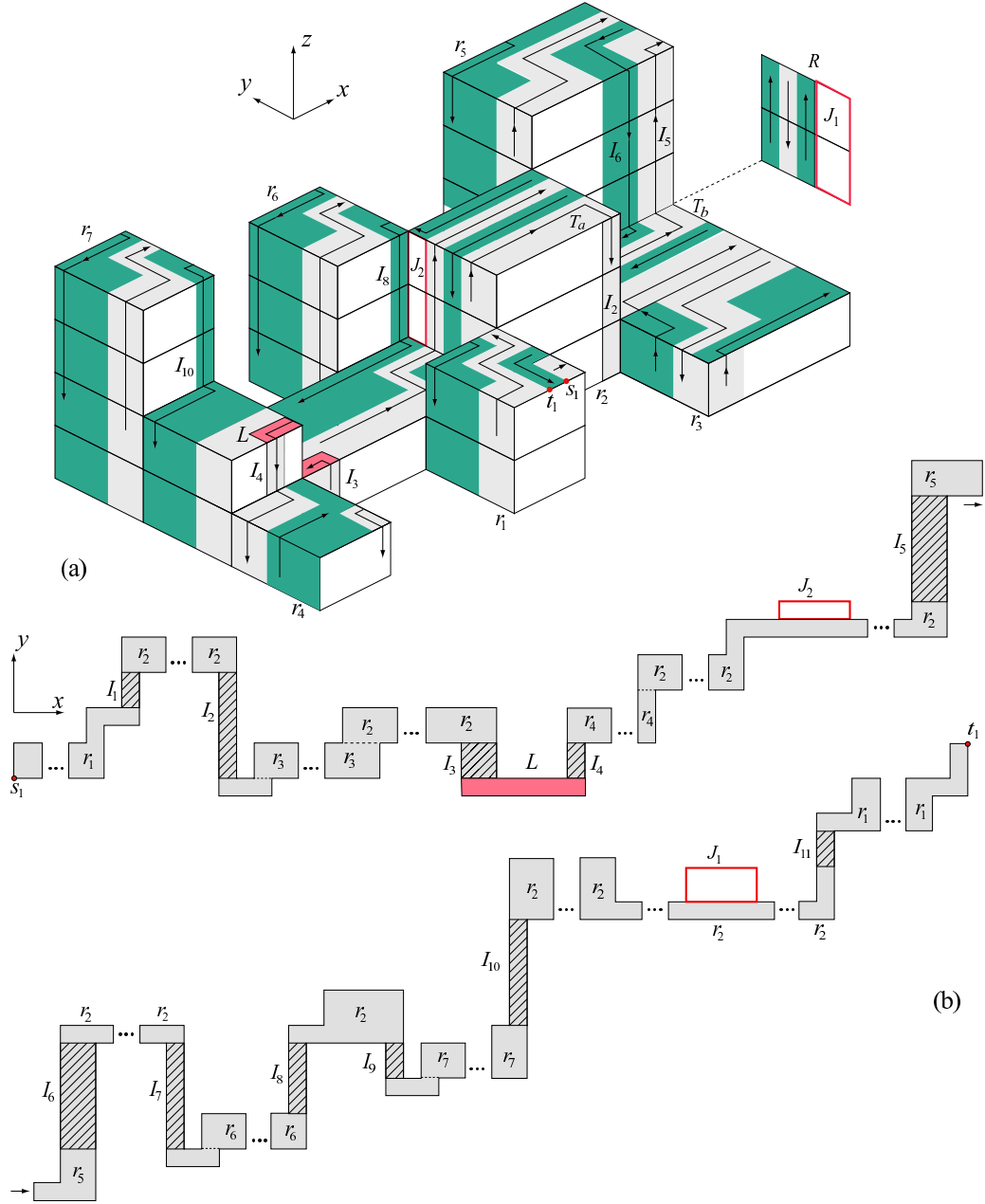}
\caption{Unfolding multiple-layer MTs. (a) Spiral $\xi_{3d}$; bottom
and back mirror views are as shown in Fig.~\ref{fig:3hshape} (b)
$\xi_{2d}$, strips $J_1$ and $J_2$ attached above; transitions
between towers are striped; piece labels correspond to MT boxes
to which they
belong.} \label{fig:3towers}
\end{figure}
%
In general, a multiple-layer MT $\P$ consists of many slabs; in this
case, we use the recursion tree for the base of $\P$ to unfold $\P$
recursively (in this sense, single-layer and multiple-layer MTs have
identical recursion structures). The recursive unfolding algorithm
is similar to the algorithm described in
Sec.~\ref{sec:base.algorithm} for single-layer MTs, with some minor
modifications to accommodate the existence of towers. In the
following we describe these modifications with the help of the MT
example from Fig.~\ref{fig:3towers}, whose base is the $3$-legged
$H$-shape single-layer MT from Fig.~\ref{fig:3hshape}.


Let $S(r')$ be the slab corresponding to a child $r'$ of $r$. When
the unfolding strip for $S(r)$ first encounters a top/bottom face
$f$ of $S(r')$ (when viewed from $z = +\infty$), the unfolding of
$S(r)$ is suspended in favor of $S(r')$. Next we discuss the two
suturing techniques used to glue the unfolding of $S(r')$ to the
unfolding of $S(r)$.

\medskip
\noindent {\bf Same-direction suture.} In this case, the bottom/top
face opposite to $f$ is used to accommodate the start unfolding
gridedge $y_{s'}$ for $S(r')$; the end unfolding gridedge $y_{t'}$
is selected as in the single-layer case.

Consider first the case when $r'$ is a front child of $r$. If
$S(r')$ is encountered while $\xi_{3d}$ runs along the top of
$S(r)$, the suture is identical to the single-layer case: a vertical
strip across the front of $S(r)$ is used to reach the bottom of
$S(r')$ (see strip $I_2$ in Fig.~\ref{fig:3towers}, reaching front
child $S(r_3)$). If $S(r')$ is encountered while $\xi_{3d}$ runs
along the bottom of $S(r)$, the suture is similar to the
single-layer case, with two simple modifications:
\begin{enumerate}
\item After using a vertical strip to reach the top of $S(r)$, a
small ``bite'' is taken out of the top of $S(r)$ to reach the top of
$S(r')$ in the single-layer case. In the multiple-layer case, it may
be necessary to extend such a bite up/down a $z$-face in order to
reach the point of the same $x$-coordinate as $y_{s'}$. This is the
case of slab $S(r_4)$ in Fig.~\ref{fig:3towers}: strip $I_3$ is used
to get from the bottom of $S(r_2)$ to the top of $S(r_2)$, after
which the ``bite'' labeled $L$ extends up a right face of $S(r_2)$
to reach the $x$-coordinate of $y_{s_4}$.

\item Unlike the single-layer case, a top bite used in the
same-direction suture is not necessarily adjacent to child $S(r')$.
In this case, a second $z$-strip  (such as $I_4$ in
Fig.~\ref{fig:3towers}) is used to reach the top of $S(r')$.
\end{enumerate}

The case in which $r'$ is a back child of $r$ is similar and is
illustrated in Fig.~\ref{fig:3towers}: strips $I_7$ and $I_{9}$
(visible in Fig.~\ref{fig:3towers}b, but not in~\ref{fig:3towers}a)
are used to make the transition from $S(r_2)$ to $S(r_6)$ and
$S(r_7)$ respectively, and strips $I_8$ and $I_{10}$ are used to
return to $S(r_2)$.

\medskip
\noindent {\bf Reverse-direction suture.} As in the same-direction
suture case, a vertical strip may be needed to make transitions
between the top of a parent $S(r)$ and the top of a child $S(r')$
that uses reverse-direction suture. This is the case for $S(r_3)$ in
Fig.~\ref{fig:3towers}, where the vertical strip $I_6$ ($I_7$) is
used to move from (to) $S(r_2)$ to (from) $S(r_5)$.

\medskip
\noindent
The result of these alterations is that $\xi_{2d}$ may
lengthen vertically, but it remains monotone in the horizontal
direction.

One final modification is necessary due to the difference in height
between towers that belong to a same slab (see for instance towers
$T_a$ and $T_b$ of $S(r_2)$ in Fig.~\ref{fig:3towers}a). In such
cases it is possible that the spiral $\xi_{3d}$ does not completely
cover the left/right faces of the slab. We resolve this problem by
thickening $\xi_{3d}$ in the $y$-direction to cover the uncovered
pieces. To be more precise, consider the vertical strip marked $J_1$
in Fig.~\ref{fig:3towers} (in the mirror view of right face $R$).
The reason $J_1$ remains uncovered is because in unfolding $S(r_3)$,
the unfolding of $S(r_2)$ suspends at the top $y$-gridedge of $J_1$
and resumes at the bottom $y$-gridedge of $J_1$. Similarly,
$\xi_{3d}$ skips over the strip marked $J_2$ in
Fig.~\ref{fig:3towers}: when the back spiral of $S(r_2)$ meets
$S(r_6)$, the unfolding of $S(r_2)$ suspends at the top $y$-gridedge
of $J_2$ and resumes at the bottom $y$-gridedge of $J_2$.

We resolve the problem of uncovered strips as follows. First, note
that every uncovered strip is on a left/right face (never a
back/front face) of a slab. This means that each left/right piece of
$\xi_{3d}$ adjacent to an uncovered strip can be thickened until it
completely covers it. This results in vertically thicker pieces in
the planar embedding $\xi_{2d}$ of $\xi_{3d}$. Because $\xi_{2d}$ is
monotonic in the horizontal direction, thickening it vertically
cannot result in overlap. It also cannot interfere with the hanging
of the front/back faces, since front/back faces attach along
horizontal ($x$-parallel) sections of $\xi_{3d}$, whereas the
thickened strips occur along otherwise unused vertical
($z$-parallel) sections of $\xi_{3d}$. Thus we have the following
result.

\begin{thm}
\label{theo:5x5} Every Manhattan Tower polyhedron can be
edge-unfolded with a $4 \times 5 \times 1$ refinement of each face
of the vertex grid.
\end{thm}

\section{Conclusion}
We have established that every $(4 \times 5 \times 1)$-refined
Manhattan Tower polyhedron may be edge-unfolded. This is the second
nontrivial class of objects known to have a refined grid-unfolding,
besides orthostacks. This is the first unfolding algorithm for
orthogonal polyhedra that uses recursion, something we believe will
be useful in developing algorithms to unfold more general shapes
that can branch in many directions.
The algorithm works on some orthogonal polyhedra that are not
Manhattan Towers, and we are working on widening its range of
applicability.

\paragraph*{Acknowledgements.}
We thank the anonymous referees for their careful reading and
insightful comments.

\bibliographystyle{alpha}
\bibliography{MT}
\end{document}